\documentclass[final]{aipproc}
\usepackage{amsmath, graphicx}
\layoutstyle{6x9}
\newcommand{\llangle}{\langle\!\langle}
\newcommand{\rrangle}{\rangle\!\rangle}

\begin{document}

\title{Non-Markovian dynamics in \\the theory of full counting statistics}



\author{Christian Flindt}{
  address={MIC -- Department of Micro and Nanotechnology,
             NanoDTU, Technical University of Denmark, Building 345east,
             DK-2800 Kongens Lyngby, Denmark}
}

\author{Alessandro Braggio}{
  address={LAMIA-INFM-CNR, Dipartimento di Fisica, Via Dodecaneso 33, I-16146, Genova, Italy}
}

\author{Tom\'a\v s Novotn\'y}{
  address={Department of Condensed Matter Physics, Faculty of Mathematics and Physics,
           Charles University, Ke Karlovu 5, 121 16 Prague, Czech Republic}
}

\begin{abstract}
We consider the theoretical description of real-time counting of
electrons tunneling through a Coulomb-blockade quantum dot using a
detector with finite bandwidth. By tracing out the quantum dot we
find that the dynamics of the detector effectively is non-Markovian.
We calculate the cumulant generating function corresponding to the
resulting non-Markovian rate equation and find that the measured
current cumulants behave significantly differently compared to those
of a Markovian transport process. Our findings provide a novel
interpretation of noise suppression found in a number of systems.
\end{abstract}

\maketitle

The theory of full counting statistics concerns the probability
$P(n,t)$ of having transferred $n$ charges through a mesoscopic
system at time $t$, when starting counting at $t=0$
\cite{Nazarov:2003}. Rather than the probability distribution
$P(n,t)$, it is often more convenient to consider the cumulant
generating function $S(\chi, t)$ defined as
\begin{equation}
e^{S(\chi,t)}\equiv  P(\chi,t)=\sum_nP(n,t)e^{in\chi},
\label{eq:cgf}
\end{equation}
from which the zero-frequency cumulants of the current can be found
in the long-$t$ limit by deriving with respect to the counting field
$\chi$ at zero, i.e.,
\begin{equation}
\llangle I^n\rrangle=\left.\frac{d}{dt}\frac{d^n
S(\chi,t)}{d(i\chi)^n} \right|_{\chi=0,\,t\rightarrow\infty},\,
n=1,2,3,\ldots
\end{equation}
In this work we consider the effects of a finite bandwidth of the
apparatus detecting charge transfers on the \emph{measured} counting
statistics. In particular, we show that the finite bandwidth makes
the effective dynamics of the detector \emph{non-Markovian}, and we
discuss how non-Markovian dynamics in general can make the counting
statistics and the corresponding current cumulants behave
significantly differently compared to Markovian transport processes.
Although, the conclusions reached below are obtained for a specific
setup, we argue that they are valid for a large class of systems.

\begin{figure}
  \includegraphics[width=0.3\textwidth, trim = 0 0 0 0,clip]{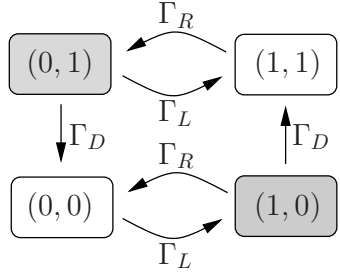}
  \caption{Detector-dot model. The system switches with rates $\Gamma_L$, $\Gamma_R$, and $\Gamma_D$
  between the states $(i,j)$, where $i=0,1$ denotes the charge state of the quantum dot and $j=0,1$
  the charge state as \emph{measured} by the detector. Each switching event between the states $(1,0)$ and $(1,1)$
  corresponds to the measurement of a single electron having entered the quantum. In the ideal
  detector limit $\Gamma_D\rightarrow\infty$, the system effectively switches between the states $(0,0)$ and $(1,1)$ with rates
  $\Gamma_L$ and $\Gamma_L$, while the other states essentially remain unoccupied. We note that similar figures can be found in Refs.\ \cite{Naaman:2006, Gustavsson:2007}.}
  \label{fig:process}
\end{figure}

We consider a model of real-time counting with a finite-bandwidth
detector \cite{Naaman:2006} recently employed in order to explain
experimental counting statistics results on electron transport
through a Coulomb-blockade quantum dot \cite{Gustavsson:2007}. In
the experiment a quantum point contact was used to monitor the two
charge states participating in transport through a nearby quantum
dot weakly coupled to source and drain electrodes and each switching
event between the two charge states was associated with an electron
either entering the quantum dot from the source electrode or leaving
it via the drain. Rather than just considering the two charge states
of the quantum dot, while keeping track of the number of electrons
$n$ that have tunneled through the quantum dot, the model also takes
into account the state of the detector that counts the electrons. In
the following $P_{ij}(n,t)$ denotes the probability that the system
at time $t$ is in a state, where the quantum dot is occupied by
$i=0,1$ extra electrons, while the detector indicates $j=0,1$ extra
electrons on the quantum dot, and $n$ electrons \emph{according to
the detector} have been transferred through the quantum dot. We
collect these four probabilities in the vector
$\mathbf{P}=(P_{00},P_{10},P_{11},P_{01})^T$ and note that
$P(n,t)=P_{00}(n,t)+P_{10}(n,t)+P_{11}(n,t)+P_{01}(n,t)$. The
counting field is now introduced via a Fourier transformation as in
Eq.\ (\ref{eq:cgf}) and the Markovian equation of motion for
$\mathbf{P}(\chi,t)$ then reads
\begin{equation}
\frac{d}{dt}\mathbf{P}(\chi,t)=\mathbf{M}(\chi)\mathbf{P}(\chi,t),
\end{equation}
where
\begin{equation}
\mathbf{M}(\chi)= \left(
  \begin{array}{cccc}
    -\Gamma_L & \Gamma_R & 0 & \Gamma_D \\
    \Gamma_L & -(\Gamma_D+\Gamma_R) & 0 & 0 \\
    0 & \Gamma_D e^{i\chi} & -\Gamma_R & \Gamma_L \\
    0 & 0 & \Gamma_R & -(\Gamma_D+\Gamma_L) \\
  \end{array}
\right). \label{eq:markov}
\end{equation}
Here, $\Gamma_L$ and $\Gamma_R$ denote the rates at which electrons
are injected and leave the quantum dot, respectively, while
$\Gamma_D$ is the rate (or the bandwidth) at which the detector
reacts to changes of the charge state of the quantum dot (see Fig.\
\ref{fig:process}). An ideal detector ($\Gamma_D\rightarrow\infty$)
is able to count every electron that is transported through the
quantum dot. On the other hand, when $\Gamma_D$ is comparable to the
electron tunneling rates $\Gamma_L$ and $\Gamma_R$, the finite
bandwidth of the detector reduces the ability of the detector to
count every electron transfer event. This, of course, affects the
measured counting statistics.

In the following, we trace out the quantum dot and show that the
resulting dynamics of the detector is non-Markovian. The quantum dot
is traced out by defining $P_j\equiv\sum_{i=0,1} P_{ij}$, $j=0,1$,
whose equations of motion read $\dot{P}_0=\Gamma_D(P_{01}-P_{10})$
and $\dot{P}_1=\Gamma_D(e^{i\chi}P_{10}-P_{01})$, respectively. The
probabilities $P_j(\chi,t)$, $j=0,1$, only contain information about
the state of the detector and the number of electrons counted by the
detector. By observing that
$\dot{P}_{10}=\Gamma_LP_{00}-(\Gamma_D+\Gamma_R)P_{10}=\Gamma_LP_{0}-(\Gamma_D+\Gamma_L+\Gamma_R)P_{10}$
and
$\dot{P}_{01}=\Gamma_RP_{11}-(\Gamma_D+\Gamma_L)P_{01}=\Gamma_RP_{1}-(\Gamma_D+\Gamma_L+\Gamma_R)P_{01}$,
we find
\begin{equation}
P_{10}(\chi,t)=\Gamma_L\int_0^{t}d\tau
e^{-(\Gamma_D+\Gamma_L+\Gamma_R)(t-\tau)}P_0(\chi,\tau)+e^{-(\Gamma_D+\Gamma_L+\Gamma_R)t}P_{10}(\chi,t=0),
\end{equation}
and a similar expression for $P_{01}(\chi,t)$. In the following, we
focus on the long-$t$ limit, where the initial condition
$P_{10}(\chi,t=0)$ (and $P_{01}(\chi,t=0)$) may safely be
neglected.\footnote{We note that the initial condition plays a
crucial role when studying finite-frequency fluctuations.} This
leads to a \emph{non-Markovian} rate-equation for
$\mathbf{p}(\chi,t)=(P_0,P_1)^T$, reading
\begin{equation}
\frac{d}{dt}\mathbf{p}(\chi,t) =
\int_{0}^{t}\mathbf{W}(\chi,t-\tau)\mathbf{p}(\chi,\tau),
\end{equation}
where
\begin{equation}
\mathbf{W}(\chi,t-\tau)=\Gamma_De^{-(\Gamma_D+\Gamma_L+\Gamma_R)(t-\tau)}
\left(
  \begin{array}{cc}
    -\Gamma_L & \Gamma_R \\
    \Gamma_L e^{i\chi} & -\Gamma_R \\
  \end{array}
\right).
\end{equation}
In Laplace space this translates to the algebraic equation
\begin{equation}
z\mathbf{p}(\chi,z)-\mathbf{p}(\chi,t=0)=\mathbf{W}(\chi,z)\mathbf{p}(\chi,z)
\end{equation}
or
\begin{equation}
\mathbf{p}(\chi,z)=\frac{1}{z-\mathbf{W}(\chi,z)}\mathbf{p}(\chi,t=0)
\end{equation}
with
\begin{equation}
\mathbf{W}(\chi,z)=D(z)\left(
  \begin{array}{cc}
    -\Gamma_L & \Gamma_R \\
    \Gamma_L e^{i\chi} & -\Gamma_R \\
  \end{array}
\right),
\end{equation}
having introduced $D(z)=\Gamma_D/(z+\Gamma_D+\Gamma_L+\Gamma_R)$. We
note that in the limit $\Gamma_D\rightarrow\infty$, $D(z)\rightarrow
1$, and the detector follows the \emph{Markovian} dynamics of the
quantum dot.

One can show (see e.g. Refs.\ \cite{Braggio:2006,Flindt:2007}) that
the cumulant generating function in the long-$t$ limit is given as
$S(\chi,t)=z^*(\chi)t$, where $z^*(\chi)$ solves the equation
\begin{equation}
z^*(\chi)-\Lambda_0[\chi,z^*(\chi)]=0.
\end{equation}
Here $\Lambda_0[\chi,z]$ is the eigenvalue of $\mathbf{W}(\chi,z)$
which for $\chi=0$ is zero, i.e., $\Lambda_0[0,z]=0$, and the
solution $z^*(\chi)$ must be chosen such that $z^*(0)=0$. We find
$\Lambda_0[\chi,z]=D(z)\lambda_0(\chi)$ with
$\lambda_0(\chi)=-(\Gamma_L+\Gamma_R)/2+\sqrt{(\Gamma_L+\Gamma_R)^2/4+\Gamma_L\Gamma_R(e^{i\chi}-1)}$,
and
\begin{equation}
z^*(\chi)=-\frac{\Gamma_D+\Gamma_L+\Gamma_R}{2}+\sqrt{\left(\frac{\Gamma_D+\Gamma_L+\Gamma_R}{2}\right)^2+\Gamma_D\lambda_0(\chi)}.
\end{equation}
For large matrices, in general, it may be non-trivial to find
$z^*(\chi)$ and more sophisticated methods, as the one we describe
in Ref.\ \cite{Flindt:2007}, may be needed. Having found the
cumulant generating function in the long-$t$ limit,
$S(\chi,t)=z^*(\chi)t$, we may calculate the current cumulants, and
here we just give the results for the first two current cumulants,
although it, in principle, is possible to obtain any cumulant having
found $S(\chi,t)$,
\begin{equation}
\begin{split}
\llangle I^1\rrangle =&\Gamma_R\left[\frac{1+a}{2}\right]\times\left[\frac{k}{1+k}\right],\\
\llangle I^2\rrangle
=&\left[\frac{1+a^2}{2}-\frac{k(1-a^2)}{2(1+k)^2}\right]\llangle
I^1\rrangle.
\end{split}
\end{equation}
These are the results also found in Ref. \cite{Gustavsson:2007}
using the model in its Markovian formulation given by Eq.\
(\ref{eq:markov}), and following that work we have also introduced
the asymmetry $a=(\Gamma_R-\Gamma_L)/(\Gamma_R+\Gamma_L)$ and the
relative bandwidth $k=\Gamma_D/(\Gamma_R+\Gamma_L)$.

It is interesting to consider the so-called Fano factor
$F\equiv\llangle I^2\rrangle/\llangle I^1\rrangle$. In the ideal
detector limit $\Gamma_D\rightarrow\infty$, we find the well-known
result $F=(1+a^2)/2$ for a \emph{Markovian} two-state model with
uni-directional transport where $1/2\leq F\leq 1$. For finite
bandwidths, the Fano factor may, however, be suppressed below $1/2$,
and for the given model, we find that the Fano factor is bounded
from below by the value 3/8 ($a=0$, $k=1$). In a number of papers,
the sensitivity of the counting statistics to coherent versus
sequential tunneling has been discussed \cite{Kiesslich:2006}, and
particularly, it has been conjectured that a suppression of the Fano
factor below $1/2$ for transport through a double barrier resonant
diode could be an indication of coherent tunneling rather than
sequential \cite{Aleshkin:2004}. The results found in the present
work show that a suppression below $1/2$ can occur due to
non-Markovian dynamics, which is not necessarily induced by quantum
coherence, but in general arises from tracing out parts of a system.
We believe that a similar interpretation can explain the recently
calculated Fano factor suppression of incoherent transport through a
single electron transistor (SET) coupled to a nano-mechanical
resonator \cite{Haupt:2006}. There, we believe that the dynamics of
the SET effectively is non-Markovian due to the coupling to the
resonator, which in turn can explain the suppression of the Fano
factor below $1/2$.

In conclusion, we have presented a study of the full counting
statistics of electron transport through a Coulomb-blockade quantum
dot as measured by a detector with finite bandwidth. In particular,
we have calculated the current cumulants of the measured charge
transport described by a non-Markovian rate equation obtained by
tracing out the quantum dot and only considering the dynamics of the
detector. Our results show that non-Markovian effects may strongly
effect the charge transport statistics.

\end{document}